\documentclass[conference]{IEEEtran}

\IEEEoverridecommandlockouts
\usepackage{cite}
\usepackage{amsmath,amssymb,amsfonts}
\usepackage{algorithmic}
\usepackage{graphicx}
\usepackage{textcomp}
\usepackage{xcolor}
\usepackage{url}

\usepackage{subcaption} 

\def\BibTeX{{\rm B\kern-.05em{\sc i\kern-.025em b}\kern-.08em
    T\kern-.1667em\lower.7ex\hbox{E}\kern-.125emX}}
\begin{document}

\title{Beyond Ray-Casting: Evaluating Controller, Free-Hand, and Virtual-Touch Modalities for Immersive Text Entry}

\author{
\IEEEauthorblockN{
Md. Tanvir Hossain,
Mohd Ruhul Ameen,
Akif Islam,
Md. Omar Faruqe,\\
Mahboob Qaosar,
A. F. M. Mahbubur Rahman,
Sanjoy Kumar Chakravarty,
M. Khademul Islam Molla
}

\IEEEauthorblockA{
Department of Computer Science and Engineering, University of Rajshahi, Bangladesh\\
mth\_cse@ru.ac.bd, ameensunny242@ru.ac.bd, iamakifislam@gmail.com, faruqe@ru.ac.bd,\\
qaosar@ru.ac.bd, mmr@ru.ac.bd, sanjoy.cse@ru.ac.bd, khademul.cse@ru.ac.bd
}
}

\maketitle

\begin{abstract}

Efficient text entry remains a primary bottleneck preventing Virtual Reality (VR) from evolving into a viable productivity platform. To address this, we conducted an empirical comparison of six physical input systems across three interaction styles---Controller-Driven, Free-Hand, and Virtual-Touch---evaluating both discrete tap-typing and continuous gesture-typing (swiping), alongside a speech-to-text (Voice) condition as a non-physical reference modality. Results from 21 participants show that the Controller-Driven Tap-Gesture Combo (CD-TGC) delivers the best productivity performance, achieving speeds 2.25$\times$ higher than the slowest system and 30\% faster than the current industry standard, while reducing error rates by up to 68\%. A clear trade-off emerged between performance and perceived usability: although controller-based gesture input led on speed and accuracy, participants rated Virtual-Touch Tap-Typing highest in subjective experience, scoring 80\% higher on the System Usability Scale (SUS) than the lowest-rated alternative. We further observe that Free-Hand interaction remains limited by tracking stability and physical fatigue, whereas Voice input introduces practical constraints related to privacy, editing control, and immersive engagement. Together, these findings characterize the tension between throughput and natural interaction in immersive text entry and provide data-driven guidance for future VR interface design.

\end{abstract}


\section{Introduction}
\label{sec:intro}

The emergence of accessible, high-fidelity hardware such as the Meta Quest 3 and premium platforms like the Apple Vision Pro marks a transition for Virtual Reality (VR) from a purely entertainment-focused medium to a platform for serious spatial computing \cite{android_central_2023}. As the developer community actively expands the frontiers of immersive software, the demand for applications capable of sustaining professional workflows is growing rapidly \cite{Aufegger2022}. However, for VR to become a truly viable environment for ``real work''—ranging from drafting correspondence to software development—it must overcome a fundamental barrier: efficient text entry. 

The current industry standard, controller-driven tap-typing via ray-casting, is notoriously slow and ergonomically taxing during medium-to-long sessions \cite{Boletsis2019, pietroszek2018raycasting}. This limitation effectively restricts VR's utility for tasks requiring sustained input. Interestingly, a potential solution may already exist in the mobile domain. Gesture typing (or swipe-typing) has revolutionized smartphone interaction, yet its application within immersive environments remains critically under-explored.

\subsection{Research Motivation and Scope}
Despite the proven efficiency of gesture typing on 2D screens, there is a notable scarcity of comprehensive studies evaluating its performance across the diverse interaction paradigms available in 6-DoF environments. While prior work has examined controller-based methods \cite{Boletsis2019} or mid-air gestures in isolation \cite{Jimenez2018}, few studies have systematically benchmarked gesture typing against traditional tap-typing across the full spectrum of modern inputs. This gap hinders the development of intuitive interfaces that could transform VR from a consumption medium into a productivity powerhouse.

To address this, our study conducts an empirical comparative analysis of two primary entry modes—discrete Tap-Typing and continuous Gesture-Typing — across three distinct interaction systems. We evaluate \textbf{Controller-Driven} methods employing controller-based ray-casting, \textbf{Free-Hand} methods leveraging the headset’s native optical hand-tracking API, and \textbf{Virtual-Touch} interfaces simulating direct manipulation of a virtual keyboard surface. Additionally, a speech-to-text (Voice) condition was evaluated as a non-physical input modality to contextualize performance differences across interaction types.
 Data was collected from 21 participants to quantify performance in terms of speed (WPM), accuracy (Error Rate), and subjective user satisfaction \cite{Brooke1996, IJsselsteijn2008}.

\subsection{Contributions}
This work contributes the following findings to the field of VR Human-Computer Interaction:

\begin{enumerate}
    \item \textbf{Performance Benchmark:} We demonstrate that the \textit{Controller-Driven Tap-Gesture Combo} is the superior method for productivity, achieving speeds 2.25 times faster than the slowest system and 30\% faster than the industry standard, with significantly reduced error rates.
    \item \textbf{Experience Insight:} We identify that \textit{Virtual-Touch Tap-Typing} offers the highest subjective usability (SUS), suggesting that the familiarity of physical touchscreens translates effectively to virtual environments despite lower raw speeds \cite{Brooke1996}.
    \item \textbf{Modal Validation:} We validate that gesture typing is a transformative method for VR, offering a more intuitive and efficient alternative to traditional point-and-click typing when paired with precise tracking.
    \item \textbf{User Factors:} We analyze the correlation between user performance and their prior experience, finding that familiarity with mobile swiping significantly predicts proficiency in VR gesture interfaces.
\end{enumerate}
\section{Related Work}
\label{sec:related_work}

The rapid evolution of consumer VR hardware has necessitated a parallel evolution in text-entry techniques. While foundational research focused on adapting 2D paradigms to 3D spaces, recent efforts have shifted toward optimizing interaction nuances for 6-DoF environments \cite{Grubert2018}. This section reviews the pivotal studies in controller-based and free-hand interaction, highlighting the specific limitations that necessitate a comparative re-evaluation.

\subsection{Controller-Based Interaction}
Handheld controllers remain the primary interface for consumer VR, functioning analogously to a mouse in desktop computing. In a comprehensive empirical review, Boletsis and Kongsvik established that controller-based ray-casting (pointing at keys) and drum-like keyboards currently represent the efficacy baseline for the field \cite{Boletsis2019}. Their work confirmed that while ray-casting is intuitive, it often suffers from the physical constraints of hunting and pecking for keys, which limits maximum input speed compared to real-world typing.

\subsection{Alternative and Free-Hand Modalities}
Beyond standard controllers, researchers have investigated alternative inputs to enhance accessibility and immersion. Yu et al. explored head-rotation as an input mechanism \cite{yu_headbased}. While this offers a hands-free alternative valuable for users with limited mobility, it imposes significant neck fatigue and lacks the precision required for high-speed productivity.

The removal of physical hardware to achieve natural interaction has been a long-standing goal in HCI. Jimenez and Schulze pioneered the evaluation of mid-air text entry using optical tracking, suggesting that bare-hand interaction offers a superior user experience in theory \cite{Jimenez2018}. However, practical implementations have historically faced severe bottlenecks. Early optical tracking struggled to interpret rapid finger movements accurately \cite{Markussen2013}, and perceptible latency often disrupted the cognitive feedback loop essential for typing \cite{voigt2020handtracking}. Furthermore, prolonged mid-air interaction requires users to hold their hands unsupported, leading to rapid shoulder fatigue, a phenomenon widely documented as \textbf{Gorilla Arm Syndrome} \cite{Hincapie-Ramos2014,shneiderman1983direct,Markussen2013}.

\subsection{Research Gap}
Despite the breadth of existing literature, two critical gaps remain that this study seeks to address. 

First, while \textbf{gesture typing} (swiping) has become the de facto standard for single-handed input on mobile devices, it remains significantly under-explored in VR. Most controller-based studies focus exclusively on discrete tap-typing, neglecting the potential of continuous stroke gestures to reduce physical strain and increase input rates.

Second, the recent release of integrated, high-fidelity hand-tracking in headsets like the Meta Quest 3 \cite{android_central_2023} challenges previous assumptions about free-hand limitations. There is a lack of recent empirical data comparing these modern hand-tracking capabilities directly against established controller baselines and novel virtual-touch paradigms in a unified study.

\section{Methodology}
\label{sec:methodology}

To evaluate the efficacy of modern VR text entry, we engineered seven distinct input systems and conducted a rigorous within-subjects empirical study. This section details the system design, our novel data collection infrastructure, the experimental apparatus, and the metrics used for evaluation.

\subsection{System Design}
We implemented six physical interaction systems derived from a matrix of three \textit{Interaction Styles} (Controller, Free-Hand, Virtual-Touch) and two \textit{Text Entry Modes} (Tap-Typing, Tap-Gesture Combo). Additionally, a Voice Input system was implemented as a non-physical comparison condition. All physical systems utilized a standard QWERTY layout to minimize learning effects associated with unfamiliar key arrangements \cite{kapco_keyboard_layouts}.

\subsubsection{Interaction Styles}
\begin{itemize}
    \item \textbf{Controller-Driven (CD):} Utilizing standard VR controllers, this method employs ray-casting to select keys from a distance \cite{pietroszek2018raycasting}. It represents the current industry standard.

    \item \textbf{Free-Hand (FH):} Using computer vision-based hand tracking (Meta Quest), users interact via ray-casting originating from their knuckles. Selection is triggered by a ``pinch'' gesture.

    \item \textbf{Virtual-Touch (VT):} This method simulates a physical touchscreen. A virtual keyboard is positioned within arm's reach, and users ``poke'' the keys directly with their index fingers.
\end{itemize}
\begin{figure*}[t]
    \centering
    
    \begin{subfigure}[b]{0.30\textwidth}
        \centering
        \includegraphics[width=\linewidth]{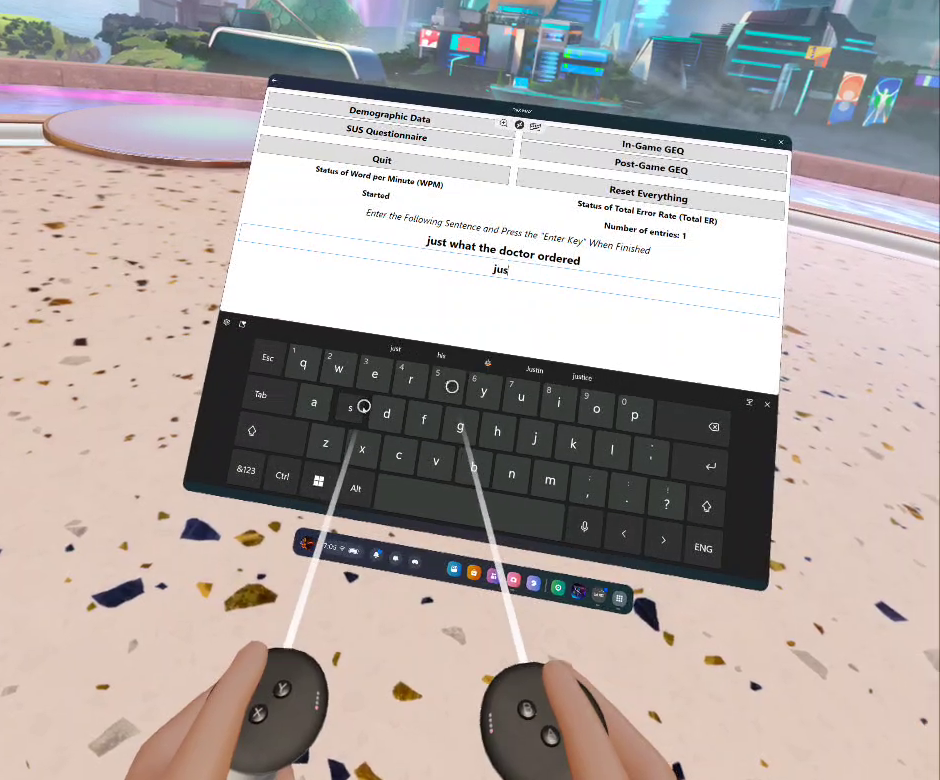} 
        \caption{Controller Tap (CD-TT)}
        \label{fig:cd_tt}
    \end{subfigure}
    \hfill
    \begin{subfigure}[b]{0.30\textwidth}
        \centering
        \includegraphics[width=\linewidth]{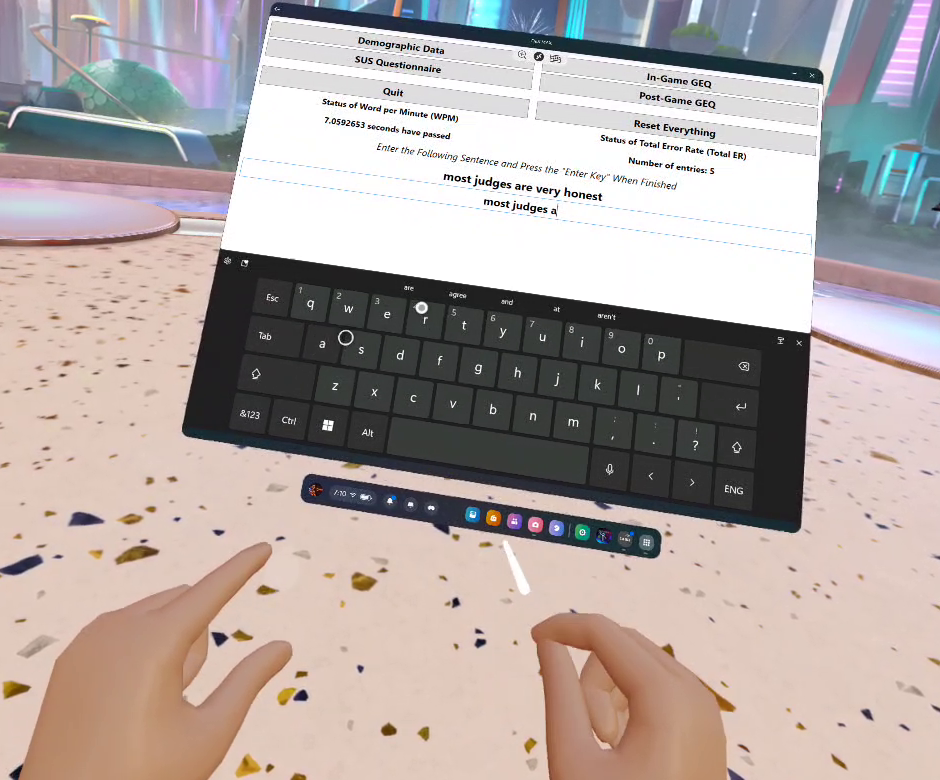} 
        \caption{Free-Hand Tap (FH-TT)}
        \label{fig:fh_tt}
    \end{subfigure}
    \hfill
    \begin{subfigure}[b]{0.30\textwidth}
        \centering
        \includegraphics[width=\linewidth]{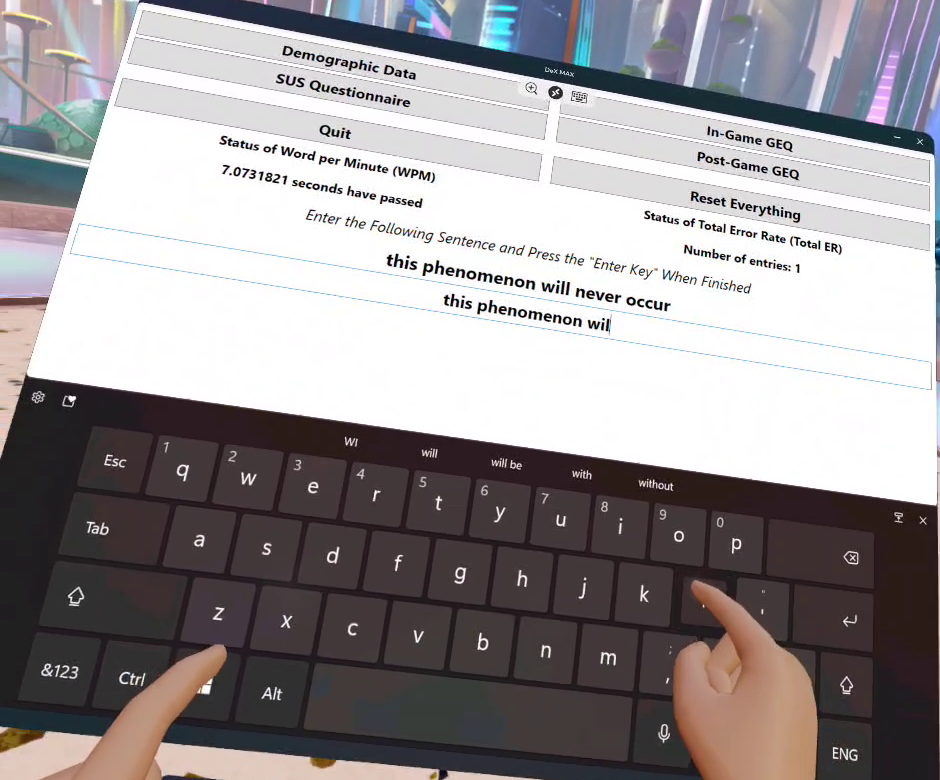} 
        \caption{Virtual-Touch Tap (VT-TT)}
        \label{fig:vt_tt}
    \end{subfigure}
    
    \vspace{1em} 
    
    \begin{subfigure}[b]{0.30\textwidth}
        \centering
        \includegraphics[width=\linewidth]{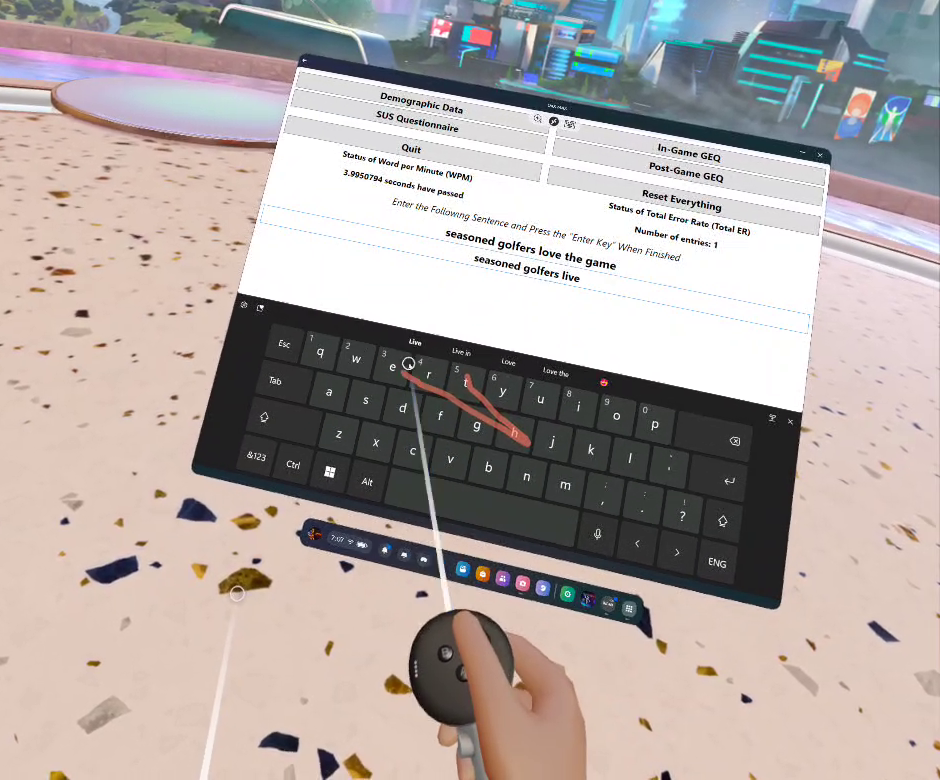} 
        \caption{Controller Gesture (CD-TGC)}
        \label{fig:cd_tgc}
    \end{subfigure}
    \hfill
    \begin{subfigure}[b]{0.30\textwidth}
        \centering
        \includegraphics[width=\linewidth]{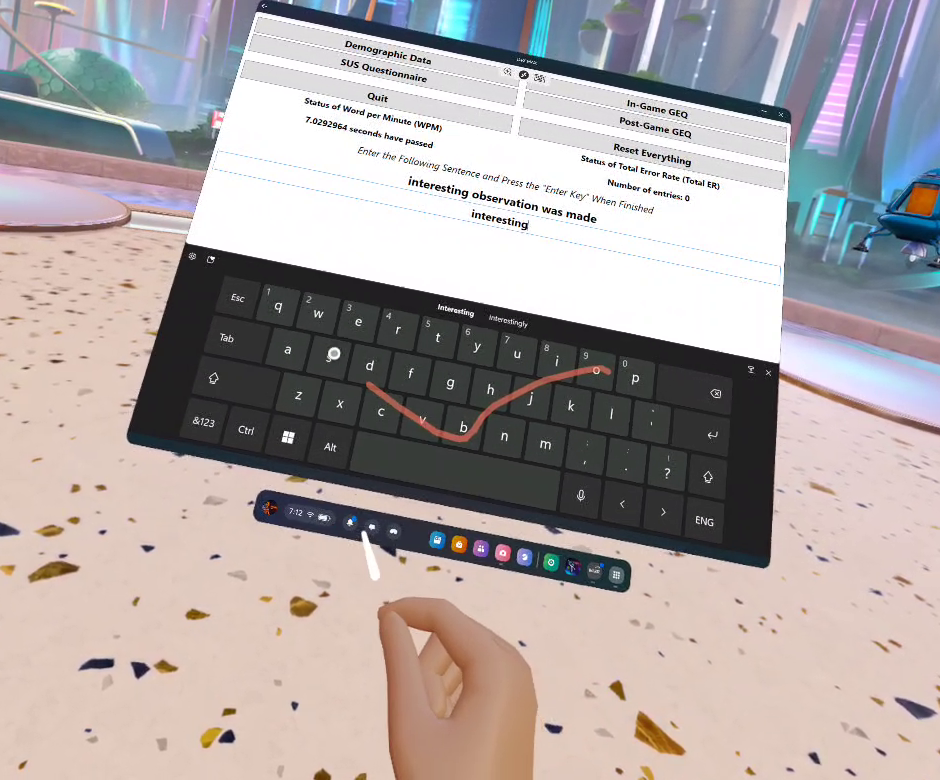} 
        \caption{Free-Hand Gesture (FH-TGC)}
        \label{fig:fh_tgc}
    \end{subfigure}
    \hfill
    \begin{subfigure}[b]{0.30\textwidth}
        \centering
        \includegraphics[width=\linewidth]{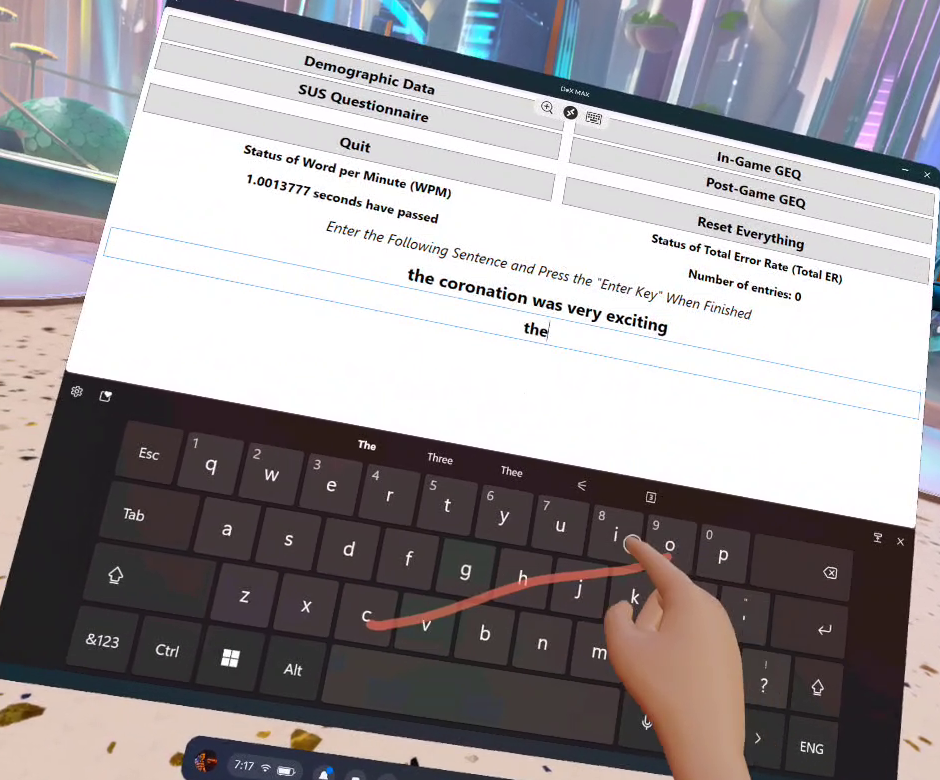} 
        \caption{Virtual-Touch Gesture (VT-TGC)}
        \label{fig:vt_tgc}
    \end{subfigure}
    
    \caption{Visual comparison of the six physical interaction systems. The top row displays the discrete Tap-Typing (TT) modes, while the bottom row displays the continuous Tap-Gesture Combo (TGC) modes where users swipe through letters.}
    \label{fig:system_grid}
\end{figure*}

\subsubsection{Text Entry Modes}
\begin{itemize}
    \item \textbf{Tap-Typing (TT):} The discrete selection of individual characters. In CD and FH styles, this is a ``point-and-shoot'' interaction. In VT, it resembles striking a physical key.
    \item \textbf{Tap-Gesture-Combo (TGC):} A hybrid mode prioritizing gesture typing (swiping). Users select the first letter of a word and drag a path through subsequent letters. Users can revert to tap-typing for corrections.
\end{itemize}

\subsection{Data Collection Infrastructure}
A persistent challenge in VR user studies is balancing the need for granular, high-frequency data logging against the limited computational headroom of standalone headsets. To resolve this, we bypassed standard standalone architectures in favor of a custom, centralized data collection ecosystem. Our design process initially explored deploying a native VR application (APK) or a custom Unity-based client-server solution. However, these approaches were ruled out; the former lacked the necessary external control for a guided experiment, while the latter introduced unacceptable development overhead regarding state synchronization.

Consequently, we engineered a centralized desktop application using C\# (WPF) to serve as the experiment's logic hub. This host application managed all computationally intensive tasks, including gesture-decoding algorithms, stimulus presentation, and real-time logging. To bridge this logic with the immersive environment, we sideloaded the Android version of \textit{Microsoft Remote Desktop} onto the Meta Quest headsets. This architectural choice effectively offloaded all processing weight to the PC, reducing the headset to a high-fidelity input/output relay. Crucially, this setup allowed us to monitor multiple headsets from a single control station with negligible latency, ensuring unified data integrity across all trials.

\begin{figure}[!t]
    \centering
    \includegraphics[width=\linewidth]{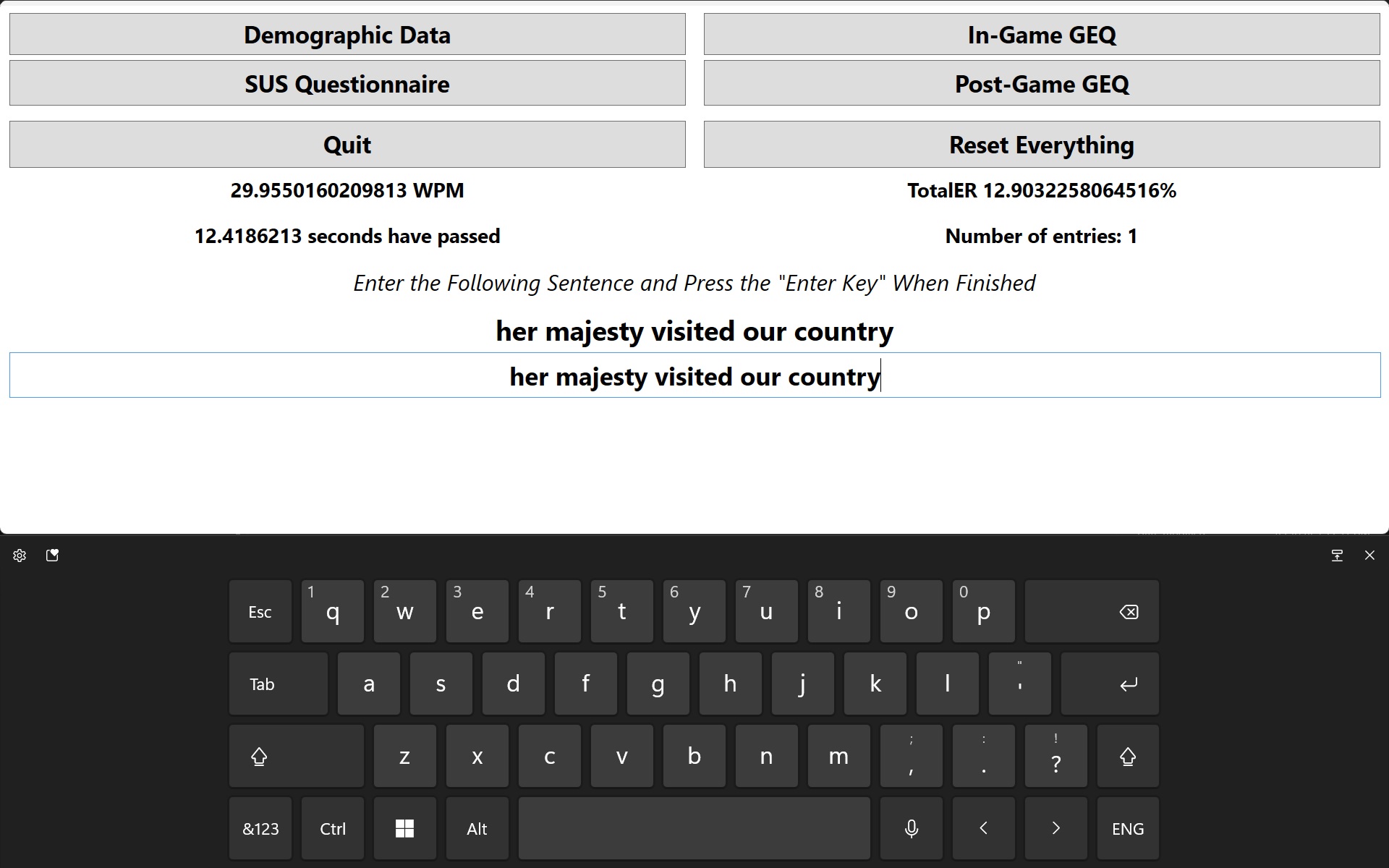}
    \caption{The custom Data Collection Application UI running on the desktop, streaming to the VR headset. It handles phrase presentation, WPM calculation, and questionnaire switching.}
    \label{fig:ui_app}
\end{figure}

\subsection{Experimental Design and Procedure}
We employed a within-subjects design where twenty-one participants ($N=21$) recruited from the Rajshahi University community tested all seven systems. To minimize order effects and fatigue, the presentation sequence was counterbalanced. The experiment utilized Meta Quest 2 and 3 headsets, with text stimuli randomly selected from the standard MacKenzie and Soukoreff phrase set \cite{soukoreff2003}. Crucially, to ensure valid testing of gesture typing algorithms, we filtered this set to exclude words shorter than three characters. Participants underwent a familiarization phase for each system before completing five measured transcription trials. In total, the study yielded \textbf{735} distinct text entry trials (21 participants $\times$ 7 systems $\times$ 5 phrases) and a comprehensive qualitative dataset comprising 147 sets of Game Experience Questionnaires (GEQ) and System Usability Scale (SUS) forms.

\subsection{Methodological Challenges and Limitations}
\label{sec:challenges}

While the experimental design aimed for rigor, two primary constraints must be acknowledged regarding the apparatus and procedure.

\subsubsection{Temporal and Functional Constraints}
The study was restricted to a single session per participant. Consequently, the results capture the "novice" experience and initial learnability rather than long-term proficiency. As noted in longitudinal studies, performance metrics and user sentiment (specifically GEQ dimensions like \textit{Tiredness} and \textit{Positive Affect}) often evolve as users overcome the initial learning curve \cite{MACKENZIE2013157}. Furthermore, to establish a raw baseline performance, no auto-correction or auto-completion algorithms were implemented. While this ensures a direct comparison of motor performance, it likely underestimates the practical WPM achievable in commercial applications where intelligent decoding is standard \cite{alharbi2020}.

\subsubsection{Infrastructural Latency}
The decision to utilize a Remote Desktop bridge—while crucial for centralized data logging and external control—introduced distinct technical challenges compared to a native VR application (APK). Although high-performance streaming was utilized, the dependency on wireless transmission inevitably introduced minor latency variations. This setup required substantial user-level customization to ensure compatibility across different headsets. While functional, this added layer of complexity may have introduced slight variability in user performance, particularly in the \textit{Free-Hand} conditions where real-time feedback loops are critical for precise interaction.

\subsection{Evaluation Metrics}
We employed standard quantitative metrics for text entry and validated questionnaires for qualitative feedback.

\subsubsection{Quantitative Metrics}
\begin{itemize}
    \item \textbf{Words Per Minute (WPM):} Speed was calculated using the standard formula \cite{MacKenzie2002}:
    \begin{equation}
        WPM = \frac{|T| - 1}{S} \times 60 \times \frac{1}{5}
    \end{equation}
    Where $|T|$ is the length of the transcribed string and $S$ is the time in seconds.
    
    \item \textbf{Total Error Rate (TER):} To account for both corrected and uncorrected errors, we used the TER metric \cite{soukoreff2003}:
    \begin{equation}
        TER = \frac{INF + IF}{C + INF + IF} \times 100\%
    \end{equation}
    Where $INF$ is Incorrect Not Fixed, $IF$ is Incorrect Fixed, and $C$ is Correct keystrokes.
\end{itemize}

\subsubsection{Qualitative Metrics}

An initial questionnaire was administered once at the beginning of the session to capture demographic information and prior experience with VR and mobile gesture typing. 

After each experimental condition, participants completed both the Game Experience Questionnaire (GEQ) \cite{IJsselsteijn2008} and the System Usability Scale (SUS) \cite{Brooke1996}, allowing for per-system subjective evaluation. 

Finally, semi-structured interviews were conducted at the end of the session to gather comparative reflections across all interaction modalities.


\section{Results and Discussion}
\label{sec:results}

\begin{figure*}[!t]
    \centering
    \begin{subfigure}[b]{0.42\textwidth}
        \centering
        \includegraphics[width=\linewidth]{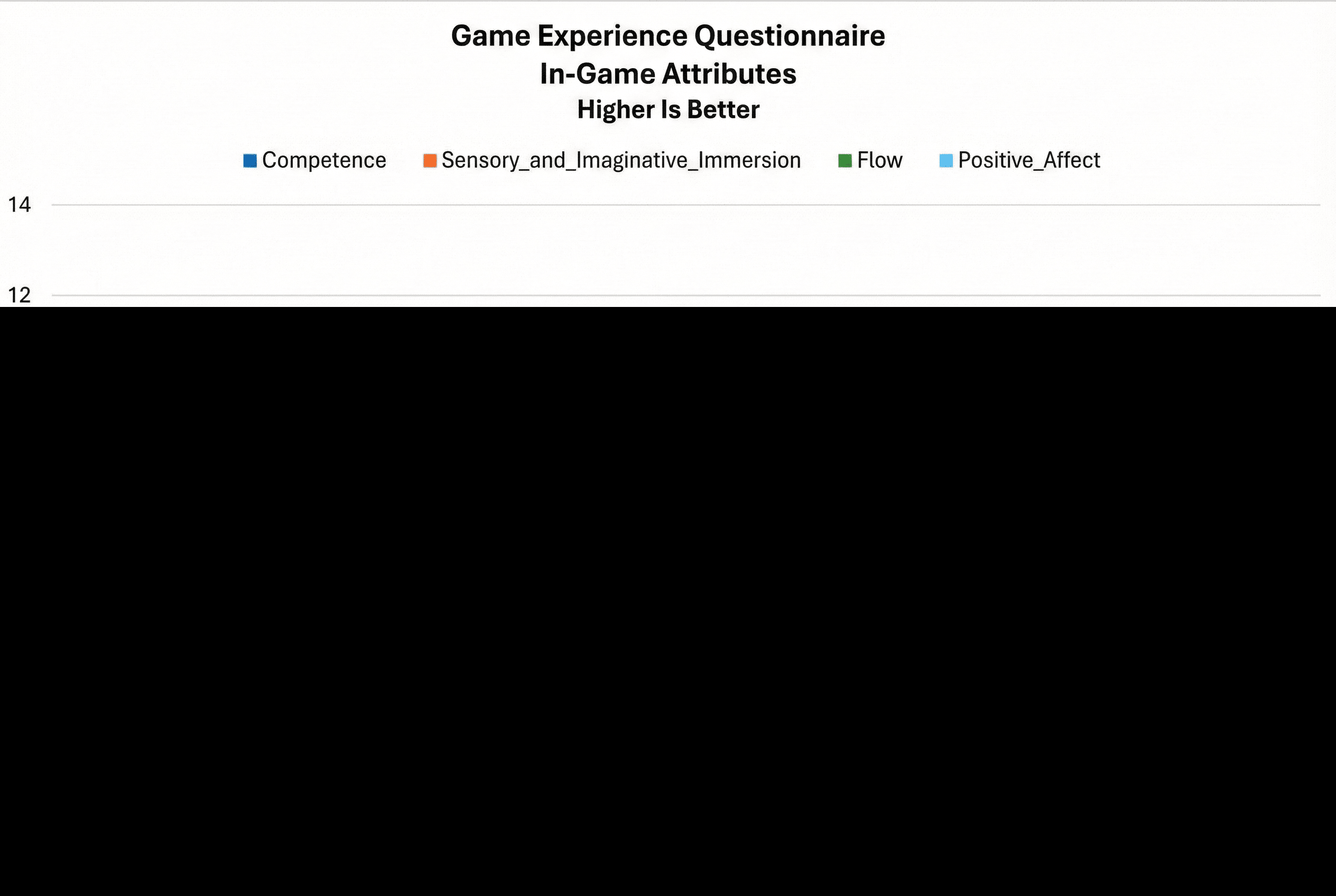}
        \caption{Positive Attributes (Higher is Better)}
        \label{fig:geq_plus}
    \end{subfigure}
    \hfill
    \begin{subfigure}[b]{0.42\textwidth}
        \centering
        \includegraphics[width=\linewidth]{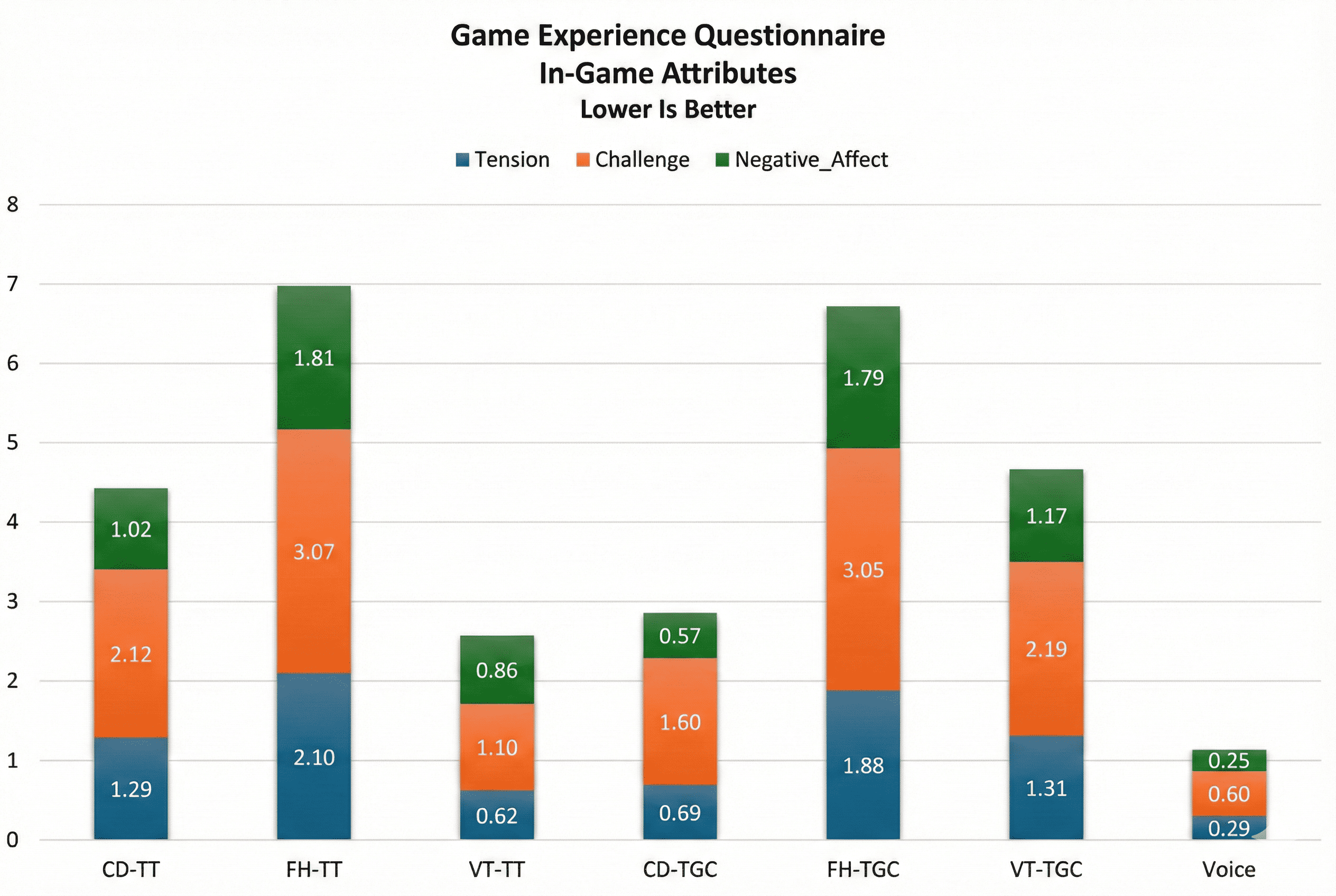}
        \caption{Negative Attributes (Lower is Better)}
        \label{fig:geq_minus}
    \end{subfigure}
    \caption{In-Game GEQ results. VT-TT and CD-TGC score highest in Competence and Flow, while Free-Hand systems suffer from high Tension and Negative Affect.}
    \label{fig:geq}
\end{figure*}

This section details the findings from our analysis of 735 text entry trials. We evaluate the systems across three primary dimensions: quantitative performance metrics (Speed and Accuracy), qualitative user experience assessments (GEQ and SUS), and subjective user preferences.

\subsection{Quantitative Performance}

\subsubsection{Speed and Accuracy}
\begin{figure}[!t]
    \centering
    \includegraphics[width=0.4\textwidth]{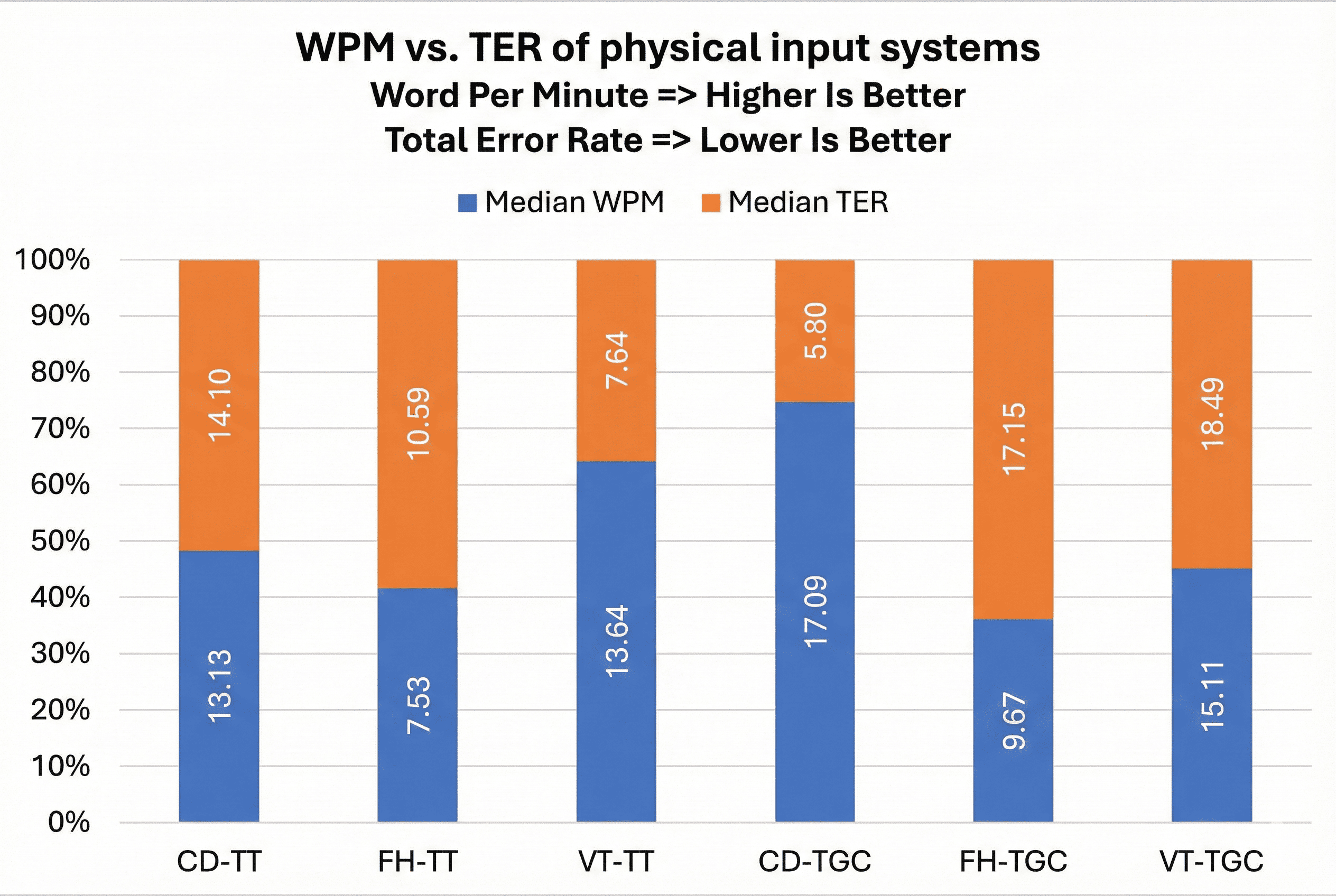}
    \caption{Median Words Per Minute (WPM) vs. Median Total Error Rate (TER) across all physical systems. CD-TGC demonstrates the optimal balance of high speed (17.09 WPM) and low error rate (5.80\%).}
    \label{fig:wpm_ter}
\end{figure}

Our analysis reveals a clear performance hierarchy among the interaction styles. As illustrated in Fig. \ref{fig:wpm_ter}, the \textbf{Controller-Driven Tap-Gesture Combo (CD-TGC)} distinguished itself as the superior physical input method. It achieved a median speed of \textbf{17.09 WPM}, outperforming the current industry standard (CD-TT) by \textbf{30\%} and surpassing the slowest system (FH-TT) by a factor of \textbf{2.25} times. Crucially, this speed did not come at the cost of precision; CD-TGC also demonstrated the highest accuracy among physical systems with a Total Error Rate (TER) of just \textbf{5.80\%}. This represents a substantial \textbf{58\% reduction in errors} compared to standard tap-typing and \textbf{68\% fewer errors} than the least accurate system (VT-TGC).

In contrast, both Free-Hand (FH) systems yielded the lowest performance metrics. The lack of tactile feedback and inherent jitter in current hand-tracking technology resulted in error rates climbing as high as 17.15\% for FH-TGC. While the Voice Input system served as a high-performance baseline with a median speed of 154.75 WPM and a low error rate of 2.12\%, its non-physical nature introduces specific situational constraints that are discussed later.

\begin{figure}[!t]
    \centering
    \includegraphics[width=0.6\linewidth]{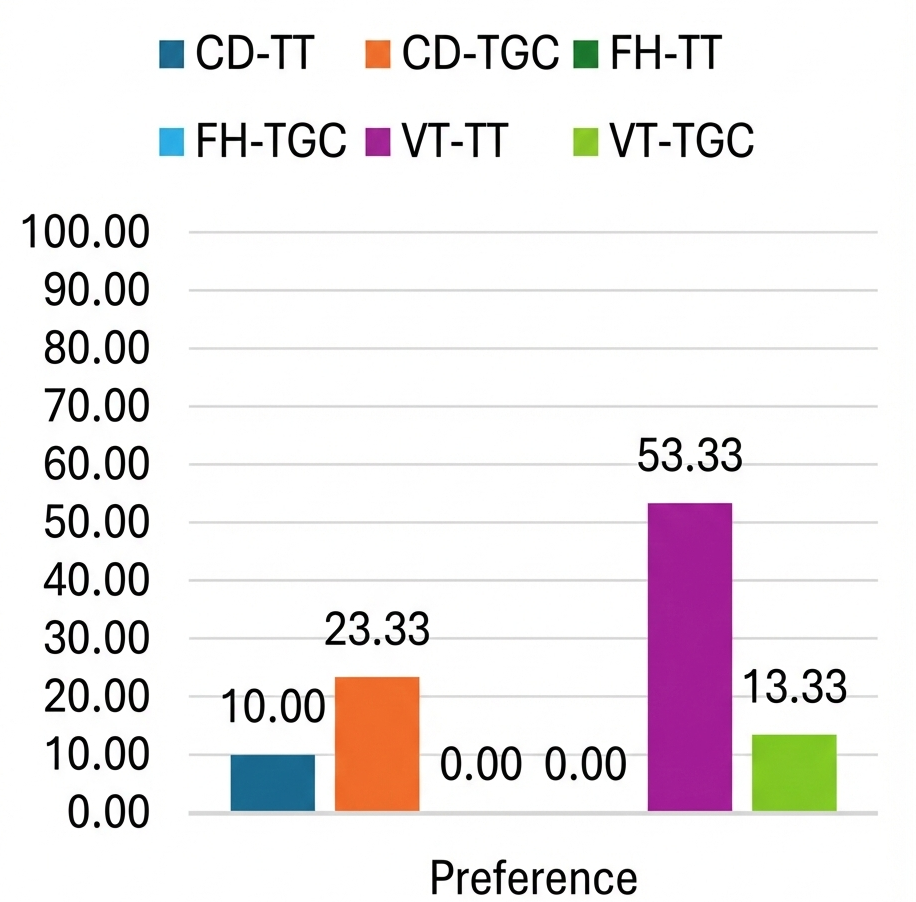}
    \caption{User preferences for Physical Systems. Virtual-Touch Tap-Typing (VT-TT) was the clear favorite among participants.}
    \label{fig:pref_phys}
\end{figure}

\begin{figure}[!t]
    \centering
    \includegraphics[width=0.5\linewidth]{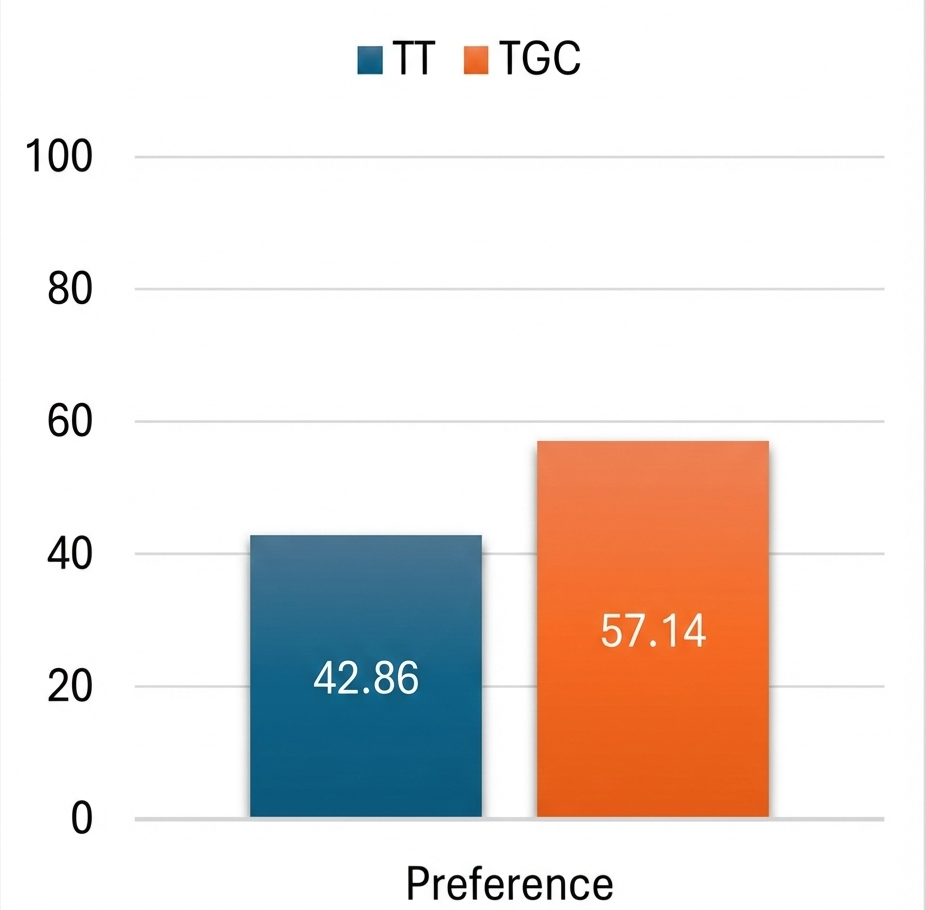}
    \caption{User preferences for Entry Mode. A majority preferred Gesture-Typing (TGC) over discrete Tap-Typing (TT).}
    \label{fig:pref_mode}
\end{figure}

\subsubsection{Impact of Experience}
We also observed a significant correlation between user demographics and performance. Participants with prior experience in mobile gesture typing performed markedly better across VR conditions. Notably, this "digital dexterity" appeared to transfer even to discrete *Tap-Typing* modes, suggesting that familiarity with 2D touch interactions provides a measurable advantage when transitioning to spatial interfaces.

\subsection{Qualitative User Experience}

While the controller-based gesture system dominated the quantitative metrics, the subjective user experience painted a different picture regarding perceived usability and comfort.

\subsubsection{System Usability Scale (SUS)}
\begin{figure}[!t]
    \centering
    \includegraphics[width=\linewidth]{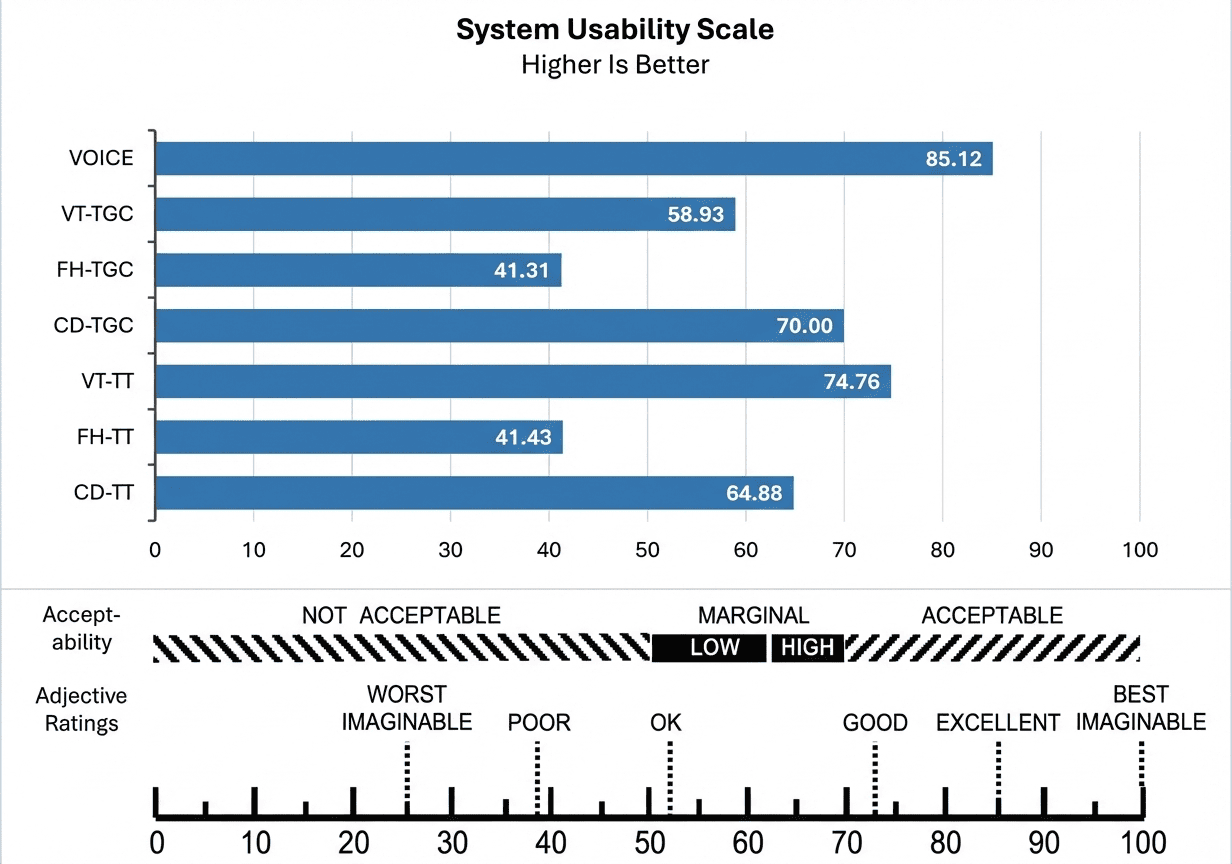}
    \caption{System Usability Scale (SUS) scores. Virtual-Touch Tap-Typing (VT-TT) is the preferred physical method, approaching the "Excellent" range.}
    \label{fig:sus}
\end{figure}

Despite being physically slower, the \textbf{Virtual-Touch Tap-Typing (VT-TT)} system received the highest subjective ratings (Fig. \ref{fig:sus}). It achieved a SUS score of \textbf{74.76}, placing it \textbf{15\% higher} than the industry standard CD-TT (64.88) and a remarkable \textbf{80\% higher} than the lowest-rated system (FH-TGC). This strongly suggests that users find the direct manipulation metaphor—simply reaching out to ``poke a virtual screen—far more intuitive and cognitively less demanding than the remote ray-casting abstraction used by controllers.

\subsubsection{Game Experience Questionnaire (GEQ)}
The GEQ results provided deeper insight into the users' internal states (Fig. \ref{fig:geq}). Participants reported the highest sense of \textbf{Competence} and \textbf{Flow} when using the VT-TT and CD-TGC systems. Interestingly, a distinct anomaly emerged with Voice Input: despite being the fastest method, it received the lowest score for \textbf{Flow} (2.19). Interviews revealed that users felt dictation was passive and disconnected from the virtual environment, breaking their sense of immersion. Conversely, Free-Hand systems scored highest on Tension and Negative Affect, with post-game data confirming that unsupported mid-air interaction caused the most physical tiredness—an effect commonly referred to as \textbf{Gorilla Arm Syndrome} \cite{Hincapie-Ramos2014,shneiderman1983direct,Markussen2013}.

\subsection{User Preference}
Subjective preferences, visualized in Figs. \ref{fig:pref_phys} and \ref{fig:pref_mode}, mirrored the usability scores rather than the raw performance data. A clear majority of users identified VT-TT as their preferred physical system, valuing its ease of use over the raw speed of the controller-gesture combo. However, when asked about entry mode specifically, the majority preferred Gesture-Typing over Tap-Typing, indicating a desire for swiping mechanics provided they are implemented on a reliable interaction platform.

\subsection{Critical Analysis of Interaction Modalities}

The contrast between performance metrics and user feedback reveals an important tension in VR interface design: what is fastest is not always what feels best. While \textbf{CD-TGC} clearly achieved the highest speed and accuracy, participants consistently preferred \textbf{VT-TT}, describing it as more familiar and easier to control. This suggests that interaction choices in immersive environments depend heavily on context. Direct, touch-like interaction appears well suited for lightweight or occasional tasks, whereas controller-based gesture input may better support sustained, productivity-focused work.

We also observed practical challenges with unsupported mid-air gestures. The weaker performance of VT-TGC compared to its controller-based counterpart points to issues such as depth perception uncertainty and the lack of tactile resistance. During rapid swiping, users often struggled to maintain stable finger trajectories, resulting in increased tracking loss and higher error rates. These findings suggest that mid-air interaction may benefit from haptic feedback or stronger spatial anchoring to improve stability and confidence.

Although Voice input achieved the highest raw speeds, participants expressed reservations about privacy, editing precision, and a reduced sense of immersion. Rather than replacing manual input, voice appears better positioned as a complementary channel—useful for high-bandwidth input, but not a full substitute for embodied interaction in immersive environments.

\section{Conclusion}
\label{sec:conclusion}

This study empirically demonstrates that the \textbf{Controller-Driven Tap-Gesture Combo (CD-TGC)} is the superior method for VR productivity, significantly outperforming alternative systems in both speed and accuracy. Conversely, \textbf{Virtual-Touch Tap-Typing (VT-TT)} emerged as the most user-friendly interface, highlighting a distinct trade-off between raw performance and intuitive interaction. While Free-Hand and Voice modalities offer theoretical benefits, they remain constrained by current tracking precision and privacy concerns. Future research should prioritize multimodal integration between manual gesture and voice input, as well as the incorporation of haptic feedback, to enhance precision and reduce physical fatigue in immersive text entry.

\bibliographystyle{IEEEtran}
\bibliography{references}

\end{document}